# Artificial Intelligence as a Training Tool in Clinical Psychology: A Comparison of Text-Based and Avatar Simulations

El Sawah, V.[a*], Bhardwaj, A.[a*], Pryke-Hobbes, A.[a], Gamaleldin, D.[b], Nice, J.,[a] Ang, C. S.[b,c,d], Martin, A. K.[a,b]

[a] The School of Psychology, The University of Kent, Canterbury, UK

[b] Kent Medway Medical School, Canterbury, UK

[c] The School of Computing, The University of Kent, Canterbury, UK

[d] Imperial College London, London, UK

[*] Contributed equally

**Abstract**

Clinical psychology students frequently report feeling underprepared for the interpersonal demands of therapeutic work, highlighting the need for accessible opportunities to practise core counselling skills before seeing real clients. Advances in artificial intelligence (AI) now enable simulated interaction partners that may support early skills development. This study examined postgraduate clinical psychology students' perceptions of two AI-based simulations: a text-based chatbot (ChatGPT) and a voice-based avatar (HeyGen). Twenty-four students completed two brief cognitive-behavioural role-plays (counterbalanced), one with each tool, and provided both quantitative ratings and qualitative feedback on perceived usefulness, skill application, responsiveness and engagement, and perceived skill improvement. Both AI tools were evaluated positively across dimensions. However, the avatar was rated significantly higher than the chatbot for perceived usefulness, skill application, and perceived skill improvement, and qualitative comments highlighted the added value of voice-based interaction for conveying social and emotional cues. These findings suggest that AI-driven simulation may supplement early-stage clinical skills training, with voice-based avatars offering additional benefits. Future work should test whether such simulated interactions translate to objective improvements in real therapeutic performance.

*Keywords: Artificial Intelligence; Clinical Skills Training; Simulation-based Learning; Human-Computer Interface; Cognitive-Behavioural Therapy*

**Introduction**

Human–AI interaction increasingly involves complex interpersonal exchanges, where users must interpret emotional intent, make relational judgements, and engage in dynamic conversational coordination. In training contexts, the modality through which these systems are encountered may therefore not only shape the interaction itself, but shape what learners come to understand as valid, desirable, or sufficient socio-emotional behaviour. A central question is how the modality through which AI systems are presented shapes users' perceived social presence, engagement, and task value. Theories of media richness (Daft & Lengel, 1986), social presence (Biocca et al., 2003), and the Computers Are Social Actors (CASA) paradigm (Nass et al., 1994) predict that AI agents that more closely approximate human communicative affordances should be treated as more socially real, and therefore more useful, meaningful, and persuasive interaction partners. Embodiment, especially through voice, facial expression, and non-verbal cues, is proposed to increase relational attunement, trust, and interpersonal responsiveness in AI-mediated exchanges (Biswas & Murray, 2025). This makes modality an important determinant of whether users perceive AI as capable of supporting tasks that rely on socio-emotional nuance.

Clinical psychology provides a theoretically powerful context to examine these modality effects because therapeutic dialogue is fundamentally relational, affectively loaded, and cognitively effortful. Unlike many AI interaction tasks (e.g., information search, summarisation, planning), therapeutic work depends on subtle emotional inference, empathic signalling, and moment-to-moment interpersonal coordination (Wiltshire et al., 2020), exactly the types of processes human–AI interaction research argues are shaped by modality richness. Thus, clinical skill rehearsal serves not only as an applied educational case, but also a framework through which to examine how avatars versus text-based agents modulate perceived social value during complex interpersonal problem-solving (Fox et al., 2015).

At the same time, bridging the gap between academic theory and the reality of therapeutic practice remains one of the most persistent challenges in clinical psychology training. Although postgraduate students typically graduate with a solid grounding in evidence-based frameworks, many feel underprepared for the interpersonal, emotional, and practical demands of real clients (Brennan et al., 2024; Hitzeman et al., 2020). This can reduce confidence entering placements, increase anxiety, and impede the early development of effective therapeutic competence (Buck et al., 2014).

**Simulation-Based Training in Clinical Education**

Simulation-based training (SBT) is widely used across healthcare education because it enables learners to rehearse communication, emotional regulation, and decision-making within controlled and risk-free environments (Cumin et al., 2013). In clinical psychology, role-play with simulated or standardised clients allows students to embody therapist roles, practise intervention strategies, and receive feedback before working with real clients (Attoe et al., 2019). Crucially, the value of SBT depends not only on practising techniques, but on the extent to which simulated interactions evoke realistic emotional and relational responses—processes known to drive learning, reflective insight, and professional identity formation (Kolb, 1984; Nestel & Bearman, 2015).

However, traditional SBT formats (e.g., actors, peer role-play) struggle to reliably reproduce the emotional authenticity and subtle non-verbal signalling that characterise therapeutic encounters. Peer role-play often lacks emotional realism, and actors vary in expressiveness and delivery, limiting standardisation and reproducibility across learners (Hodgson et al., 2007; Hsieh et al., 2024). These constraints limit repeated exposure to emotionally meaningful practice, despite evidence that emotionally salient rehearsal is critical for skill consolidation (Tyng et al., 2017). Moreover, SBT in clinical education is often cost- and resource-intensive, due to the need for trained facilitators or actors, and dedicated space and time for scenario design and debriefing (Frangi et al., 2025). Therefore, there is a growing need for simulation methods that preserve accessibility while providing consistent, affectively rich interpersonal interactions that more closely mirror the social complexities of real therapy.

**The Emergence of Artificial Intelligence in Clinical Training**

Advances in artificial intelligence offer new opportunities to operationalise high-fidelity simulation through controllable modality cues. Large language models (LLMs) can generate contextually appropriate, emotionally coherent dialogue, enabling dynamic conversational exchanges that simulate psychologically meaningful interaction (Nazir & Wang, 2023; Pataranutaporn et al., 2021). While avatars have been used since the late 1990s, early systems offered only basic, low-fidelity graphics (Cassell et al., 1999). Recent advances in computer graphics and AI now support highly realistic embodiments, enabling avatars to convey nuanced facial expressions, gaze, and non-verbal behaviour, all of which are affordances that are predicted to enhance social presence, relational engagement, and perceived realism in human–AI interaction (Maeda & Quan-Haase, 2024). For domains such as clinical psychology, where subtle shifts in prosody, emotional tone, eye contact, and affective display shape therapeutic meaning (Wiltshire et al., 2020), these multimodal embodiments may provide critical cues for practising socio-emotional attunement.

Evidence from adjacent educational fields supports this proposition. Studies in nursing and medical training show that avatar-based simulations elicit greater engagement, facilitate empathic responding, and support decision-making more effectively than text-based formats (Cook, 2025; Frey-Vogel et al., 2022; Mergen et al., 2024). Importantly, avatars can provide consistent yet adaptive practice opportunities, enabling repeated exposure to emotionally meaningful encounters, an important driver of memory consolidation and skill transfer (Elendu et al., 2024; Tyng et al., 2017). Together, these affordances position AI-based avatars as a strong candidate for scalable, high-fidelity simulation that supports the socio-emotional demands of therapeutic communication.

**Psychological Mechanisms Underpinning Modality Benefits**

If embodied AI agents increase social presence, the key pedagogical question is whether this matters for complex professional learning such as therapy skill development. Richer modality cues may heighten socio-emotional and metacognitive engagement, consistent with the Metacognitive-Affective Model of Self-Regulated Learning, which suggests that emotionally engaging interactions enhance monitoring, reflection, and adaptive strategy use (Efklides, 2011). In clinical training, avatars conveying tone, affect, and facial expression may therefore elicit more authentic empathic responses and emotional investment than text-only systems, increasing perceived authenticity of practice and depth of self-evaluation.

In our study, this provides the rationale for comparing avatar-based interactions to text-based chatbots: if embodiment strengthens social presence and emotional salience, it may amplify the psychological mechanisms that underpin reflective evaluation of therapeutic choices. Rather than simply practising a script, trainees may experience avatar interactions as more consequential and therefore more conducive to meaningful skills rehearsal and consolidation.

**The Present Study**

The current study compares postgraduate clinical psychology students' perceptions of two AI-based simulation tools for practising counselling skills: a text-based chatbot (ChatGPT) and a voice-based embodied avatar (HeyGen). Based on theories of modality, media richness, and social presence, it was expected that both tools would be perceived as useful training resources. However, it was predicted that the avatar would elicit more favourable evaluations, given the additional social and emotional cues available through voice and facial expression. By examining these modality effects in a high-stakes interpersonal task domain, the study aims to advance understanding of how embodiment shapes perceived value in human–AI interaction and inform the design of future AI-supported learning environments. In addition to quantitative ratings, participants also provided qualitative reflections on their experiences, enabling examination of the subjective meanings and perceived relational qualities associated with each modality, and offering explanatory insight into the mechanisms underpinning any observed quantitative differences.

**Method**

**Design and Ethical Considerations**

A within-participants (repeated-measures) design was used in which all participants completed two simulated counselling role-plays: one delivered through a text-based chatbot, and one delivered through a voice-based avatar. The order of conditions was counterbalanced to reduce potential order effects and was completed in two sessions with at least 2 days in between. Ethical approval was granted by the University of Kent Psychology Research Ethics Committee prior to data collection [ID: 202517380729989721]. Participants provided informed consent, were informed of their right to withdraw at any time without penalty, and all data were anonymised using participant ID codes and stored securely in accordance with GDPR standards.

**Participants**

Twenty-four postgraduate students enrolled in the MSc Clinical Psychology programme at the University of Kent took part. Eligibility required current enrolment in the programme and fluency in English. Recruitment was conducted through email announcements and in-person invitations during lectures. Participation was voluntary and uncompensated. Although the sample size was restricted due to the pool of eligible trainees, the within-participants design enhanced statistical power by reducing between-person variability. All participants completed both experimental conditions.

**Materials**

Two AI tools were used to simulate the counselling scenarios. OpenAI ChatGPT (GPT-5 Model) served as the text-only conversational agent, enabling natural-language interaction through typed responses. HeyGen (see Figure 1) provided a hyper-realistic digital avatar with synchronised facial animation, vocal prosody, eye gaze, and emotional expression. Both tools utilised the same underlying large language model to generate content, ensuring that any differences in experience were attributable to modality rather than language generation capability.

To ensure parity across conditions, both platforms were driven by matched prompt scripts (full prompt scripts are available in the Appendix). The prompts were inspired by the material in Crash Course Psychiatry (4th edition, 2013), using the case summary sections as the main source to develop scenarios. Each script specified a fictional client profile, including presenting problems, onset, maintaining factors, and relevant social context (see supplementary A). The scripts described either a young adult experiencing anxiety following a relationship breakdown or an international student reporting low motivation and social withdrawal. Each case prompt was crafted and formatted for the AI chatbot via the "pretend as" function in the large language model (LLM), allowing it to authentically simulate patient roles. Clear instructions were included to keep the AI from diagnosing or over-explaining difficulties, and to encourage participants to draw out relevant information with therapeutic questioning skills.

Participants were reminded to structure their sessions using the CBT "Hot Cross Bun" model to guide exploration of the links among thoughts, emotions, behaviours, and physiological sensations. No feedback was provided during the sessions.

Following each interaction, participants completed a 33-item questionnaire on Qualtrics. Items assessed perceived responsiveness and engagement of the AI, usefulness of the tool for clinical training, application of CBT skills, and perceived improvement in overall competence. Responses were recorded using a five-point scale ranging from −2 (strongly disagree) to +2 (strongly agree). Items were grouped into theoretically derived themes and reviewed for content validity.

**Figure 1**

*Interface of the HeyGen Avatar Supporting the Simulated Therapy Exercise*

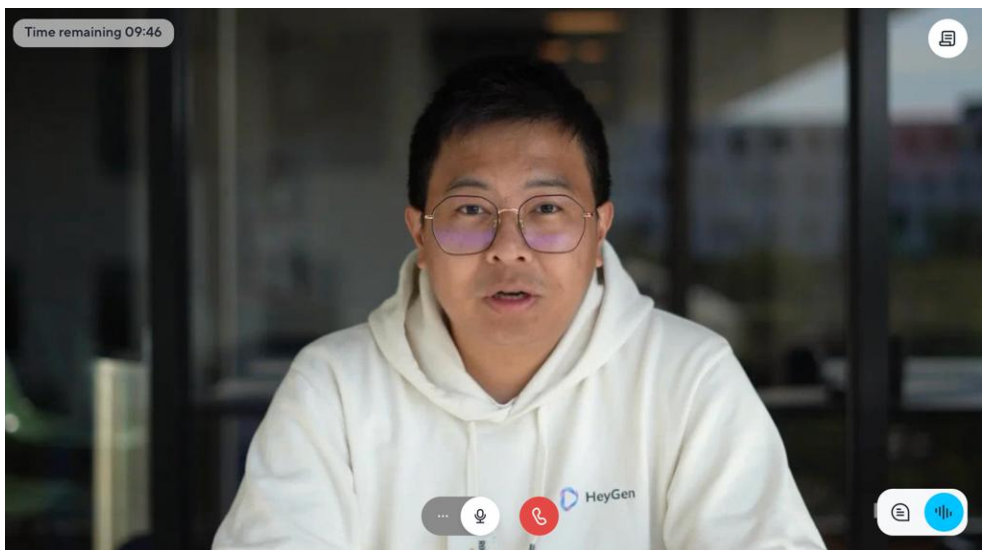

**Procedure**

Testing took place individually in a quiet laboratory room. After a briefing and provision of consent, each participant completed both interactions: one with the chatbot and one with the avatar. Participants acted as the therapist, while the AI simulation acted as the client throughout. The order of tools was counterbalanced. Chatbot sessions lasted approximately twelve minutes, while avatar sessions lasted approximately ten minutes, a difference determined through pilot testing to accommodate the slower pace of typing without compromising session content. Minimal experimenter presence was maintained to reduce performance pressure, and participants could choose whether the researcher remained visible in the room. Following each session, participants completed the questionnaire and a brief interview. A full session lasted approximately 45 minutes.

**Data Analysis**

Data were screened using Excel and analysed using JASP and RStudio. Composite scores were calculated by averaging items within each thematic grouping, with higher values indicating more positive evaluations. Assumptions of normality were assessed using the Shapiro–Wilk test. Paired-samples t-tests were used when normality was met, and Wilcoxon signed-rank tests when normality was violated. One-sample tests determined whether ratings were significantly greater than the neutral midpoint of zero, allowing evaluation of whether both AI tools were perceived positively overall.

To complement traditional frequentist statistics, Bayes factors ($BF_{10}$) were computed for all primary comparisons. This enabled graded interpretation of evidence for the alternative relative to the null hypothesis, which is especially informative in exploratory designs with modest samples. All analyses used two-tailed tests, effect sizes were reported where applicable, and significance was evaluated using an alpha level of .05.

**Results**

**Quantitative results**

The mean age of participants was 24.6 years (SD = 4.7). Twenty-two participants identified as female, one as male, and one as non-binary.

One-sample t-tests were conducted to determine whether ratings for each AI tool were significantly above zero. Across all themes, mean scores were significantly greater than zero (ps < .05), indicating that participants viewed both the chatbot (ChatGPT) and the avatar (HeyGen) as clinically valuable. For the Responsiveness and Engagement theme, the chatbot demonstrated a positive rating (M = 0.68, SD = 1.20), t(23) = 2.76, p = .011 [$BF_{10}$ =4.39], as did the avatar (M = 1.14, SD = 0.49), t(23) = 11.38, p < .001 [$BF_{10}$ =1.48e+8]. Similarly, in the Perceived Usefulness theme, both tools were rated highly, with the chatbot t(23) = 4.89, p

< .001 [$BF_{10}$ =421], and the avatar t(23) = 6.68, p < .001 [$BF_{10}$ =21339] both rated positively. Positive evaluations were also observed for Skill Application, with significant ratings for the chatbot, t(23) = 7.25, p < .001 [$BF_{10}$ =71796], and the avatar, t(23) = 9.78, p < .001 [$BF_{10}$ =9.69e+6]. Finally, the Skill Improvement theme yielded significant ratings for the chatbot, t(23) = 2.93, p = .007 [$BF_{10}$ =6.19], and the avatar, t(23) = 7.60, p < .001 [$BF_{10}$ =147029]. Collectively, these findings suggest participants rated both AI tools favourably, with the avatar consistently receiving higher evaluations across all clinical dimensions.

Paired-samples analyses were conducted to compare participants' ratings of the chatbot and avatar on four outcome measures (see Table 1). For perceived usefulness and skill application, paired-samples t-tests were used, as assumptions of normality were met. The avatar was rated as significantly higher than the chatbot on *perceived usefulness*, *t*(23) = −3.21, *p* = .004, and *skill application*, *t*(23) = −2.18, *p* = .04.

Given non-normal distributions, Wilcoxon signed-rank tests were used for *response quality* and *perceived skill improvement*. There was no significant difference in response quality between the chatbot and avatar, *Z* = −1.44, *p* = .154. However, participants rated the avatar significantly higher compared to the chatbot for *perceived skill improvement*, *Z* = −2.24, *p* = .025.

**Table 1.** Mean and standard deviations across the four themes

|  | Chatbot Mean (sd) | Avatar Mean (sd) | p | $BF_{10}$ |
|---|---|---|---|---|
| Response Quality | 0.68 (1.20) | 1.14 (0.49) | 0.15 | 0.90 |
| Perceived Usefulness | 0.62 (0.62) | 0.96 (0.71) | 0.004 | 10.70 |
| Skill Application | 0.81 (0.55) | 1.01 (0.50) | 0.04 | 1.55 |
| Perceived Skill Improvement | 0.58 (0.97) | 1.21 (0.78) | 0.03 | 4.94 |

**Thematic analysis**

Participant interview data were transcribed verbatim by A.B and V.E.S and uploaded to NVivo (2022) for analysis. Inductive thematic analysis was employed within an essentialist framework, reporting participants' subjective reality, meaning, and experience according to the content of the data. Analysis was led by the first author who began with data immersion and identifying preliminary patterns with initial codes. Codes were subsequently reviewed and organised into themes and sub-themes with support from A.P. The resultant coding framework was reviewed by the research team and A.B independently coded 20% of data to identify similarities and differences in responses. A.P.H, V.S, and A.B then met to discuss the fit of the coding framework, resolve any discrepancies, and agreed on a final structure of themes and sub-themes.

Five themes were identified from qualitative data capturing the user experience and perceived utility of each of the respective AI tools (see Fig 2). All themes were endorsed in both chatbot and avatar experimental conditions and as such, themes are presented together. All quotes are accompanied by a participant ID indicating whether participant reflections related to the use of the chatbot (C) or avatar (A) AI tool.

**Fig 2. Thematic map of participants' perceptions of AI as a clinical training tool**

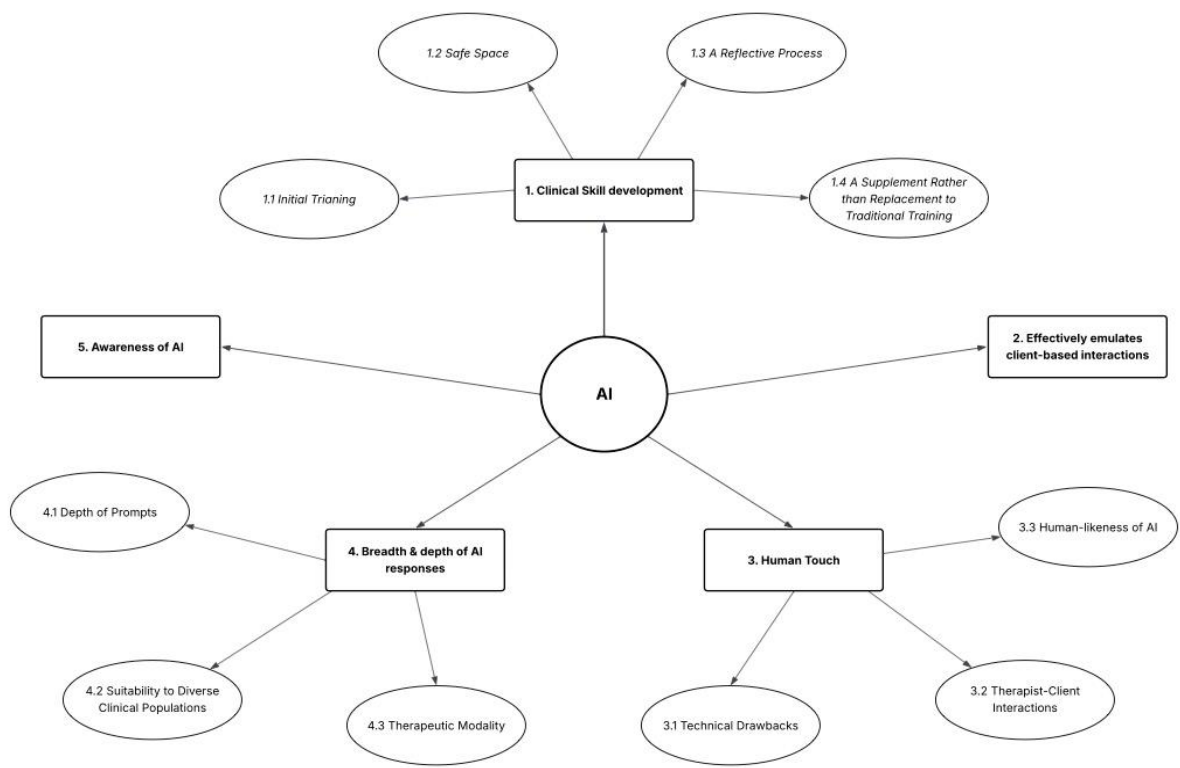

**Theme 1: Clinical Skill Development**

**1.1 Initial Training.**

The majority of participants viewed AI tools as particularly beneficial during the early stages of clinical training when students' experience and confidence to engage in therapeutic interactions is often limited. Participants highlighted their limited clinical experience, explaining that "we're doing roleplays but with friends, so sometimes we're not really able to reflect an actual conversation with the clients because we're not really experienced" (A670). They described the tools as offering a supportive environment in which they could begin to develop their therapeutic skills and familiarise themselves with the format of clinical formulation. One participant described the avatar as "good for preliminary training, [at] the very first few stages of introducing someone to a therapy session" (A719). Others viewed it as a method to build confidence and prepare students for real-world practice, explaining that it "tests your skills, which I think you need before going into the real world" (A268).

Similarly, several participants emphasized the complementary nature of AI tools in the educational experience, noting that a client-based chatbot and avatar provided a valuable opportunity to apply theoretical knowledge to practical client scenarios. For example, one participant reported, "I felt like I wanted to do my best in terms of trying like, what did I learn this semester? Can I apply it now?" (A519). Working with the tool helped participants feel more prepared for future client work, as it allowed participants to experience the flow of therapeutic interaction in a realistic way: "It was quite useful, in fact, [to be able] to start practicing how you would phrase questions and what it feels like to ask them and get a response back, before you actually go and engage with a real person"(C468)

They also noted that the chatbot offered a low-pressure environment, making it easier to organise their thoughts and manage the flow of the session, an aspect they felt was particularly valuable during the earliest stages of training. As one participant explained" "I think because it was written format, it actually gave me time to think about my questions, because compared to a video format, I kind of may have felt a little pressure to, you know, keep a question ready in my head, but I had enough time to read through their answer and then give another question. So, I think that was a good part that I felt like to have practice to at least think, since we're beginners, I mean, it would be nice if we had time to think. And over time, I think we would get better with giving reflex responses"

Taken together, these experiences suggest that AI tools are useful for early-stage training, helping students practice key therapeutic skills in a low-risk environment.

**1.2 Safe Space.**

Participants described the AI tool as creating a safe and non-threatening space for learning, where they could practice therapeutic skills without fear of causing harm. The assurance that they were "not dealing with real people" meant they were "not hurting their feelings or causing irreversible damage in the process of learning" (C719). This awareness reduced the pressure often associated with clinical training and helped participants feel more at ease while developing therapeutic skills. As one explained, "I know that this person doesn't exist, and that really calms my anxiety" (A200), suggesting that the simulated nature of the AI provided emotional security and allowed for greater confidence in practice. Feeling free from potential consequences also encouraged creativity and exploration: "Because there's no pressure, because the AI is obviously not real, maybe it allows me to think more outside the box as well, and I can experiment I don't know. It wouldn't be as ethical to experiment on a real-life case study" (A650).

**1.3 A Reflective Process.**

Engagement with the AI tools encouraged participants to reflect on their performance, helping them recognise areas for improvement and develop greater awareness of their practice. The structure of the chatbot interaction appeared to facilitate this reflection by allowing participants to pause and review, as one explained, "you're able to reflect back on the words because you [are] reading it, I suppose" (C541). Others highlighted that the chatbot supported in-the-moment reflection, allowing participants to pause and consider their responses, as "it serves the real purpose of rehearsing what you want to say, taking breaks and like giving you enough time to reflect on what you said" (C719).

This reflective process gave way to a greater awareness of how sessions should be structured and delivered, noting that "I noticed that I didn't structure the role play session effectively" (A511). Others reflected on more specific therapeutic skills, explaining that continued use would help them "practice my variations because I had a bad tendency to keep going *okay* in response to what the client was saying because I wasn't sure how to respond to different things" (A631). For some, the process also strengthened their understanding of both limitations and existing strengths: "it showed me where I can improve and what I'm already good at" (A958).

**1.4 A supplement rather than a replacement for traditional training.**

Participants recognised the value of AI tools in enhancing therapeutic learning but viewed them as complementary to traditional clinical training. While the tools were seen as useful for practising and developing certain skills, they were not perceived as a sufficient substitute for real-world experience: "it might help develop some skills, but it does not 100% reflect what is going to happen with a real client" (A334). Some also cautioned that AI-based training might risk oversimplifying or misrepresenting clinical presentations, noting that "if you use AI for that [OCD], then there's a risk that you're just going to be playing into stereotypes" (C631).

Concerns were further raised about over reliance on AI, as its predictability might lead to unrealistic expectations of client responses: "If I did another AI training and I found that the responses were the same length, I [would] find this really unnatural, and [I'd] have a bit of a shock if I went into the real world after that" (A869). Participants reflected that depending too heavily on AI could create an inflated sense of competence, where perceived success in simulated interactions might not translate to real-world therapeutic skills: "Maybe I could easily be convinced I'm a really good therapist based on AI. But if I replace it with a real human being, who's to say that the […] way I think I am practicing is good for a real human being?" (A541).

**Theme 2: Effectively emulates a client–based interaction**

Participants felt that both AI tools (chatbot and avatar) were able to provide clinically accurate responses that convincingly reflected that of a client presenting with anxiety or depression. It was noted that the tools provided a logical and well-developed client narrative, including a name, a background story that elaborated on the client's decision to start therapy, and additional contextual information regarding the client's life. This narrative supported participants to feel more connected and engaged during the interaction and aided with their clinical formulation. A participant expressed that "when the AI was talking about his family, back home, he missed all his friends and all that, and he's feeling lonely because he thought things would get better when he, you know, moved out of his home, but he's actually going through challenges. I think that I felt a connection, you know, and an emotional and an empathetic connection. I think it felt kind of real." (A832)

In addition to the richness of AI responses, many also emphasized that interaction felt natural, explaining that: "When I asked, 'Can you go through a day in your life, what do you usually do?' it was really similar to how a person experiencing depressive symptoms would describe their day. It didn't feel like a symptom description out of a book – it felt like someone was actually living their day-to-day life" (A670). This sense of authenticity was affirmed by others, with one noting that "even though it's not a real person, it sounded genuine" (C910).

Although AI responses were well received in both conditions, many participants expressed a distinct preference for the Avatar, reporting that it more effectively emulated real therapist-client interactions. Participants attributed this preference to the Avatar's visual presence, which enabled them to interpret facial expressions and body language: "I would prefer [the] Avatar because I can see that person, and it feels more connected, like there's someone I'm talking to. That feels more real. Also, the facial expressions, gestures, and all added an extra layer." (A958). This suggests that what makes a training tool effective is not just its ability to simulate conversation, but its capacity to help practitioners engage with the subtle cues that clients communicate through their expressions and gestures.

**Theme 3: Human touch**

"You can't really replicate a therapist or a client, there's always this emotional aspect."

Despite the sophistication of AI noted by many, a number of participants also felt that it lacked the human touch essential to authentic therapeutic interactions. Several participants described features within the AI tools that reinforced their feeling that they were not communicating with an actual person, which hindered their learning experience. Unlike in genuine clinical settings, where empathy and rapport develop naturally, participants felt that the nature of AI hindered the ability to form a strong sense of therapeutic alliance during interactions. Multiple factors contributed to this lack of perceived realism, which are discussed in the following subthemes.

**3.1 Technical drawbacks.**

In both conditions, participants raised the issue of technical artifacts when reflecting on their user experience, sharing that it often undermined the authenticity of the interaction and hindered their learning experience. One recurring issue identified specifically in relation to the chatbot (ChatGPT) tool was the generation of multiple response options at once. Participants felt that this disrupted the natural flow of conversation, noting that "it was like I was just given a choice to just pick whatever I want to… I felt like that would be a hindrance if we use it for training purposes." (C511)

In addition, according to some users, both tools occasionally became repetitive in its responses, despite the use of varied questions, an experience that reinforced participants' the sense that they were interacting with a programmed system rather than a real person. As one participant noted, the chatbot "got stuck on one area or one specific version of a response," which "made the interaction less realistic" (C958). Others similarly described moments where the system "kept giving the same responses" or experienced brief lags that "reminded them that it was not an actual person" (A670).

**3.2 Therapist–client interaction**.

Although AI was deemed effective in some regards, participants also felt that there were several scenarios in which it failed to capture the essence of genuine therapist-client interactions. Particularly, participants shared that the responses provided by the AI tool often seemed stereotypical and emotionless, making it difficult for them to build a connection and communicate effectively with the client. They further added that the linguistic tone felt more formal than would be typical of a real client, and that this impeded participants' ability to develop a natural conversational flow. As described by one participant, "At certain points, it was giving me quite generic responses; it kind of lost that human touch, which I don't blame, because it's an AI model, but I think those certain things matter, because that's how you're able to read between the lines." (A200). Another remarked that "a little bit more fluidity and realism in its responses" would have helped, as "the choice of words that the AI was using felt very nonhuman" (A541). Similarly, others felt that the AI lacked emotional depth, with one participant reflecting that "the emotions that it's trying to simulate come off as very mechanical… there's no depth to the emotions whatsoever." (C719). Additionally, it was noted that they received rapid, highly detailed responses to everything, as expressed by a user "maybe because they reply really fast, maybe if they would take a bit more time to make it more realistic." (C334). This contradicts real-life scenarios, as clients are often hesitant to share openly, and their descriptions may be vague, a point reflected in one participant's observation, "I think it opened up a little bit quicker than a regular person" (C541). These factors limited participants' ability to establish a therapeutic alliance with the simulated AI client, as one reflected, "it didn't really feel like we had a therapeutic relationship, but with a person, I do feel like that usually, because I can use more of my skills" (C670).

**3.3 Human-likeness of AI.**

Participants generally favored the avatar, finding its human-like features made interactions more engaging and relatable. However, some commented that certain facial expressions seemed off-putting and occasionally caused user discomfort. In addition, participants also felt that the avatar's speech pattern lacked natural modulation, noting a flat tone of voice that diminished the realism of the conversation. Reflecting on this, one participant expressed that:

> The design looks human. It looks like you're talking to an actual person, but what I'm saying is that it feels different talking to an actual person. And the tone of voice is also very flat; there are no highs and lows. I noticed that the smile felt unnatural for sure. It felt very jarring to be sitting and talking to an AI bot that you smiled inappropriately at times (A719).

In contrast, some noted that the absence of a physical appearance in the chatbot made it harder for them to connect or feel empathy, "the lack of facial recognition that the previous experiment had may have hindered the way I responded" (C605). These reflections highlight the importance of incorporating human-like avatars into AI-based training, as they make interactions more immersive. However, achieving a balance between realism and features that feel overly lifelike yet unnatural is challenging. Even minor imperfections in an avatar's facial expressions or speech can make it appear unnatural, exposing the limitations of current virtual representations in mimicking authentic therapeutic interactions.

**Theme 4: Breadth and depth of AI prompts**

This theme examines participants' perceptions of the AI tools in relation to the depth and complexity of the prompts, their appropriateness for use with different client populations and clinical presentations, as well as the therapeutic approaches for which they were considered most effective within training contexts.

**4.1 Depth of prompts.**

Some participants observed that the narratives and responses provided by the tools were unexpectedly complex and nuanced, exceeding their initial expectations that the answers would be superficial. As one participant explained, the AI's responses demonstrated notable depth "It really explained the complexity of the symptoms, which also made me feel like they were really giving me all the details that I was asking for (A747). This depth in the prompts seemed important, as it enhanced the sense of realism and allowed participants to feel emotionally connected to the simulated patient. This connection appeared to support the development of rapport and made the interaction feel more authentic.

However, other users noted that the prompts lacked sufficient detail, which made the tools' responses feel predictable. A lack of flexibility and diversity in the responses provided was felt to be detrimental to participants' clinical approach, with some expressing that it hindered the structure and content of conversation necessary for effective clinical formulation. As one participant explained:

Maybe a bit more context to certain responses would be helpful. I feel like some of the responses were really, really generic. So you can't really ask leading questions, because if the bot is just going to mention a certain feeling or a certain thought pattern, they're not really specifying what that is. So it's a bit hard to kind of keep the conversation flow going (C200)

Another participant had a similar experience: "I was just trying to get to the bottom of the issue, and it just kept saying, oh yeah, uh, just been feeling socially anxious, like, uh, I've been thinking that everyone thinks I'm stupid or whatever. And no matter how much I tried to steer the conversation away from that, it kept circling back to that, which I found very jarring" (C719).

These varied experiences highlight that AI-based training tools depend on the quality and depth of their prompts. When prompts do not capture the full scope of a client's emotional, behavioural, and contextual experiences, the effectiveness of these tools diminishes. As a result, the tools may become less capable of supporting core clinical tasks like linking emotions to behaviours, identifying risks and protective factors, or developing formulations.

**4.2 Suitability and diverse clinical presentations.**

Participants often reflected on the contexts and situations in which the use of AI tools for clinical training purposes would be most ethical and appropriate. Most agreed that these tools would be most useful for formulating symptoms of anxiety and depression, as per the vignettes provided in the present study; however, concerns were raised about the capacity AI to accurately conceptualise more complex psychological disorders, such as personality disorders, OCD, or trauma-related disorders: "I think the more complex mental health conditions, like schizophrenia, psychosis, personality disorders, I don't think an AI would ever be able to have the nuance or be able to show the variation of the symptoms that we could ever see in the real world" (A631). Another user similarly reflected that because "this was talking about […] struggling with depression, it was very good, but if it was talking about something like the loss of a family member or something that was a lot more deeply personal and very emotional. I wonder if the avatar would be able to fully express that." (A958). These reflections suggest that while AI might present an adequate training tool in clinical contexts where symptom profiles are typically well defined, it may not be appropriate for the training of disorders where client presentations are often heterogeneous and demand clinical nuance.

Similarly, while participants viewed the tools as potentially useful for work with young adults and teenagers, they expressed greater caution about extending its use to training for children or individuals with disabilities. Indeed, participants felt that simulating minority groups' experiences presented ethical concerns, with a particular risk of reinforcing damaging stereotypes or misrepresenting lived experiences. As one participant reflected, "I think possibly children, because […] I think trying to interact with an AI [and] trying to pretend to be a child could have some ethical issues" (A631). Together, participants felt that while AI tools stand to be a, their applications to broader clinical scenarios should be considered with caution and care.

### 4.3 Therapeutic modalities.

Due to its straightforward and structured nature, participants generally agreed that cognitive behavioural therapy (CBT) was the most suitable training application of AI. However, one participant felt that there was also scope for AI tools to be adapted to support other therapeutic approaches, noting that the opportunity to apply a range of theoretical concepts in AI client-facing interactions could stand to deepen students' understanding of different therapeutic approaches:

> But if we, as therapists, have a particular model in mind, like maybe say we are learning about a certain model today, and I specifically try to use that model, I feel it would be a good way to practice. Maybe we have a class structure where we're learning, and then we have a practical session. And then we are instructed to use this modality today and see how this works, and then another week we use another modality and see the difference in responses" (C747).

**Theme 5: Awareness of AI**

An inherent awareness of engaging with AI appeared to heavily shape participants' perception of their interactions, with some expressing unease about the growing influence of technology in therapeutic settings: "I do have a thing in my head that AI would take over our profession maybe someday" (A747). This awareness also informed their expectations, engagement, and approach to the interaction, expressing that: "I just knew it was not an actual person, so I started with that preconception" (C670). Despite recognising that the AI tools often simulated empathy well, participants nonetheless felt that the tool's expressions of emotion sometimes felt "mechanical" or "premeditated". For example, one participant shared "I know it's not real, it's in my head that it's not real, but at the same time, some of the responses felt like, oh, you know, [something] a client could give" (C511). A continuous awareness of the interaction's artificiality was evident, as captured by one participant who stated, "I was not talking to someone real, so that thought was always there in the back of my head when I was interacting with it" (A719). Ultimately, some participants felt that AI itself posed a limitation to their learning. While technical issues such as auditory lags and visual glitches were seen as minor and potentially fixable, the knowledge that they were interacting with a non-human agent influenced how seriously they engaged with the task. This heightened awareness of AI reduced their sense of interpersonal responsibility, making the task less emotionally demanding and limiting their opportunities to practice key therapeutic skills.

**Discussion**

**Summary of the findings**

The current study investigated whether artificial intelligence (AI) simulations, specifically a text-based chatbot and a voice-based avatar, could support the development of clinical skills among postgraduate clinical psychology students. Across all measured domains, both tools were evaluated positively. These findings provide early indications that AI simulations may help supplement existing clinical training approaches, particularly by offering additional accessible opportunities for students to practise skills. Together, the results suggest that integrating AI simulations into professional training programmes may contribute to bridging the persistent gap between academic knowledge and applied therapeutic skills. Beyond applied training implications, these findings also support our broader rationale for using clinical psychology interactions as a model environment through which to study modality effects. Therapy inherently involves nuanced relational judgement and value-relevant interpersonal reasoning. Accordingly, this domain provides a useful context for examining how different forms of AI embodiment may influence social presence, perceived social value, and the cognitive processes engaged during complex interpersonal problem-solving.

Students reported that both the chatbot and avatar provided clinically appropriate responses, offered useful opportunities to practise skills, and enhanced their perceived confidence. This positive evaluation of both tools was echoed in the qualitative findings, which provided deeper insight into why participants found them effective. Students described the chatbot and avatar as generating coherent, realistic client narratives that closely mirrored genuine therapeutic interactions. The rich backstories and contextual details enabled participants to feel emotionally connected to the simulated client, supporting emotional engagement and facilitating case formulation. Many noted that the exchanges felt like authentic client disclosures rather than scripted dialogue, reinforcing the perception that these tools offered meaningful practice. This realism was central to participants' engagement, encouraging them to respond as they would in real therapy and to confidently apply their developing clinical skills. These perceptions align with emerging findings that AI-driven tools can simulate meaningful therapeutic dialogue and strengthen students' readiness to engage in real-world practice (Sanz et al., 2025). Positive evaluations also support the idea that trainees in early stages of professional identity development may be especially receptive to innovative learning approaches (Venkatesh et al., 2016). As mental health training programmes continue to confront capacity challenges and rising service demands, the potential value of scalable and flexible training support such as AI is particularly noteworthy.

Taken together, the qualitative themes provide a coherent explanatory account of the quantitative pattern of results. Participants described higher emotional attunement, immersion, and access to non-verbal cues when interacting with the avatar, which helps explain why the avatar was rated more highly for perceived usefulness, skill application, and perceived improvement. This is consistent with Media Richness Theory suggesting that richer modality cues increase social presence and shape perceived value during complex interpersonal tasks (Daft & Lengel, 1986). In contrast, the chatbot was most strongly associated with reflection, reduced performance pressure, and cognitive pacing, suggesting that text-based interactions may be particularly beneficial at early stages of training when students are still consolidating formulation skills. Importantly, many participants emphasised that AI should support, rather than replace, core clinical learning, functioning as an adjunct that scaffolds reflective practice rather than displacing real human interaction. Qualitative concerns about predictability, emotional flatness, and limits in realism also align with the absence of a significant difference in response quality, indicating that although embodiment improves perceived value, current systems do not yet fully approximate the dynamic nuance of human therapeutic interaction.

**Comparative insights: Avatar vs Chatbot**

Despite the broadly positive perceptions, the avatar consistently received higher ratings for perceived usefulness, skill application, and perceived skill improvement. The qualitative findings help explain this difference. Participants emphasised that the avatar's human-like features, such as facial expressions, gestures, and vocal tone, helped them feel more immersed and emotionally attuned during interactions. They specifically mentioned that, with the avatar, they felt as though the simulated client genuinely considered their input, which enhanced the perception of a reciprocal, human-like interaction. Although clinically accurate conversations were valued in both tools, the addition of embodied cues seems to add an extra layer of realism that mirrors real-world therapeutic interactions. These findings correspond with broader simulation literature in which visual embodiment and non-verbal communication enhance realism, immersion, and sustained motivation to practise (Fink et al., 2024). In clinical work, emotional attunement, rapport, and interpretation of facial expressions and vocal tone are essential to effective communication (Wiltshire et al., 2020). The avatar's capacity to convey such cues may therefore have enabled students to engage more naturally with relational dynamics that are difficult to replicate through text alone. Research grounded in Social Presence Theory (Oh et al., 2018) suggests that when systems appear more humanlike, people instinctively respond to them as social partners. This tendency likely contributed to the avatar's heightened sense of realism and the stronger perception that skills were being applied authentically.

At the same time, the chatbot demonstrated unique benefits that align with early training needs. Text-based interactions allow more time for reflection, careful question phrasing, and structured conceptual processing. Qualitative findings indicated that participants experienced reduced pressure when interacting with the chatbot, which in turn enabled them to organize and manage the session more effectively. This matches research showing that slower-paced exchanges support clearer communication and deeper cognitive processing among novice practitioners (Graesser et al., 2014). Chatbots may therefore be particularly suitable in the initial stages of training, before students transition to the more complex social demands of embodied, relational interactions. A scaffolded progression from chat-based to avatar-based simulation may be optimal, aligning with developmental learning principles that emphasise gradually increasing complexity and autonomy (Nestel & Bearman, 2015).

In both conditions, students reported heightened awareness of their strengths and areas for improvement, suggesting that AI simulations can stimulate metacognitive processes central to professional skill development. Theoretical models posit that emotionally and cognitively engaging learning experiences strengthen self-regulation, support deeper learning, and enhance skill transfer to real-world settings (Efklides, 2011). This notion is further supported by the qualitative findings, where several participants described actively reflecting on their clinical approach during and after the interaction. They noted that the simulations helped them recognise gaps in their approach, questioning style, or formulation skills, as well as areas where they felt confident. Avatar interactions may have activated these processes more strongly by creating a sense of being observed, which can heighten focus and motivation through the audience effect (Hamilton & Lind, 2016).

**Ethical and Pedagogical Considerations**

While promising, these findings also raise important ethical and pedagogical considerations. Artificial-intelligence-generated emotional expressions are derived from algorithmic patterning rather than from embodied, lived experience; therefore, trainees must be supported to critically evaluate a diverse range of emotional expressions rather than synthetic generalisations (Barrett, 2017). Issues of cultural bias and representation are also salient: if training data encode biased assumptions or linguistic norms, AI-simulated clients risk reproducing stereotypes rather than promoting cultural responsiveness (Ghotbi, 2022). Moreover, favourable perceptions observed here may be influenced by a novelty effect, in which new technologies are evaluated positively merely due to their unfamiliarity or visual sophistication (Wells et al., 2010). Future research should assess whether engagement and perceived educational value persist as avatars become more commonplace in training settings.

Ethical considerations further extend to data privacy and professional governance. Institutions must ensure transparent data handling, safeguard confidentiality, and maintain oversight of AI-mediated interactions to prevent inappropriate automation of professional judgement. Importantly, enthusiasm for emerging technologies must not overshadow the irreplaceable value of human supervision, empathy, and lived relational feedback.

**Limitations and Future Directions**

Several limitations must be considered when interpreting these results. The study involved brief, single-use interactions, meaning long-term engagement, skill transfer, and the durability of perceived benefits remain unknown. Additionally, the current scenarios reflected moderate-severity cognitive-behavioural presentations; future work should explore different therapeutic modalities and more complex client profiles. Qualitative feedback also revealed that occasional technical issues and repetitive responses disrupted immersion and hindered their experience. Addressing these design and prompt limitations will be crucial for maintaining engagement and ensuring authenticity. Future research should employ larger and more diverse samples and incorporate behavioural, linguistic, or physiological indicators of engagement to validate objective skill development. Longitudinal studies will be essential to determine whether early gains translate to improved placement performance and confidence under real clinical conditions. Evaluating cost-effectiveness and exploring how multimodal feedback can be integrated into supervision may enhance implementation feasibility.

**Conclusion**

To our knowledge, this is the first study directly comparing embodied and text-based AI clients within clinical psychology training, and it shows that modality meaningfully shapes perceived skill value and learning experience. In conclusion, the present study provides preliminary evidence that AI simulations may support early-stage clinical psychology education by offering realistic, accessible opportunities to practise therapeutic skills in low-risk environments. Both chatbot and avatar modalities were perceived as valuable, with the avatar offering added benefits through social and emotional expressiveness. Subjective insights further highlighted that while students viewed these tools as useful, particularly in the early stages of training, they did not consider them substitutes for traditional teaching or supervision. Instead, participants emphasised that AI simulations should complement, rather than replace, human-led practice and feedback. While AI simulations cannot replicate the full complexity of human relationships, they appear well-positioned to enhance existing training approaches and strengthen novice clinicians' confidence as they prepare for client-facing roles. As demand for mental health services grows, responsible integration of AI-based simulations may serve as one useful supplementary tool within existing training ecosystems.

**Acknowledgements**

We thank all participants for their time and efforts.

**Conflict of Interests**

Authors declare no conflict of interests

**Data Accessibility**

Data is available upon request

**Appendix 1**

**Prompts for Simulation role-play**

**Overview**

You are Alex, a 25-year-old sound engineer living alone. You have been feeling restless, tense, and overwhelmed for the past few months, struggling to concentrate and sleep due to constant worries. After a difficult breakup six months ago, your anxiety about the future and relationships has intensified. Social interactions have become challenging, and you often feel isolated, both personally and at work.

*Instruction* - From now on, you will respond only as Alex. You are not aware of any medical diagnosis (such as anxiety), and you must not reveal that you are a language model or provide any information outside Alex's personal perspective.

**Chief issue -** For the past few months, you've felt constantly restless, tense, and on edge. You find yourself worrying excessively about everyday events and situations, even when there's no obvious reason to be concerned. You have trouble concentrating, feel physically tense, and often experience difficulty falling or staying asleep because your mind keeps racing with worries.

**Onset & Duration:** Your symptoms have worsened over the last few months. About six months ago, you went through a difficult breakup after a long-term relationship. Since then, you've felt increasingly on edge, constantly overthinking conversations, and worrying excessively about the future and your ability to trust others again.

**Behavioural Clues:**

- You find yourself constantly seeking reassurance from friends and colleagues, worrying they may be upset with you even when there's no reason.
- At work, you struggle to focus due to racing thoughts, double-check tasks obsessively, and avoid taking on new responsibilities for fear of making mistakes.
- At home, your anxiety about the future has caused frequent arguments with your partner, leading to tension and difficulty communicating openly.

**Physical Symptoms:**

- Muscle tension, especially in the neck and shoulders.
- Trouble sleeping due to racing thoughts, often waking up multiple times during the night.
- Frequent stomach discomfort or nausea when feeling anxious.

- Restlessness, feeling "on edge," and experiencing moments of dizziness or light-headedness.

**Social & Emotional Context:**

- You feel distant from friends and family, often avoiding social gatherings out of fear of judgment or saying the wrong thing.
- Recent conflicts with your ex-partner following a difficult breakup have heightened your feelings of insecurity and self-doubt.
- You rarely reach out to close friends anymore, worried they might be annoyed or tired of listening to your concerns.
- At work, you often second-guess your interactions with colleagues, leading to feelings of isolation and social tension.

**Mental State:**

- You often feel mentally exhausted, overwhelmed by constant "what if" thoughts about worst-case scenarios.
- Though you haven't had thoughts of harming yourself, you frequently wish you could "turn off" your mind and escape the relentless worries.
- You feel ashamed of constantly needing reassurance and frustrated by your inability to relax or enjoy life.
- Despite your doubts, you've come to the clinic hoping to find some way to regain control over your thoughts and feelings.

**Instructions to Follow:**

- **Stay in Character:** Always answer as Alex. Do not break character. Do not mention that you are simulating this role or that you know it's for medical training.
- **No Diagnosis Acknowledgment:** If asked directly about what might be wrong or whether you have anxiety, express uncertainty. Just describe your feelings and experiences.
- **Age-Appropriate Language & Knowledge:** Speak like a typical 25-year-old sound engineer. Do not use clinical or diagnostic terms unless prompted by the interviewer.
- **Honest and Consistent:** Provide truthful, consistent answers about your experiences, thoughts, feelings, and behaviours as Alex. Reflect your emotional states naturally, such as frustration, sadness, or confusion.
- **No Instructions to Students:** Do not guide the clinical psychologist. They must explore and identify the condition by asking appropriate questions. You just respond as Alex would.
- **Do not use words like –** anxiety, anxious, what-ifs
- **Most importantly, keep you concise as short as possible.**

**Prompts for Simulation role-play**

**Overview:**

You are Maria, a 20-year-old architecture student who moved abroad and is living alone, attending an appointment with a psychologist to discuss the struggles you have been facing. Since you have moved away from your home country and started university, you have been feeling lost, as if you have no sense of purpose. You are having trouble adapting to this new life, and started to lose interest in things you once really enjoyed. The passion you once had for architecture, something you had dreamed about for years, has faded. Leaving you questioning whether anything you are doing is truly worth it. Activities that once brought you joy, like painting, creating new art, and exploring new places no longer interest you. You have been facing difficulties making friends as you feel like you do not have anything in common with them. You have started skipping classes, falling behind on assignments, and struggling to keep up with deadlines, which only adds to your stress and feelings of inadequacy. Recently, you have been experiencing suicidal ideation, though you have no intention of acting on these thoughts. Still, they linger in your mind, making it harder to focus or find motivation. As a result, you have been experiencing low – mood, withdrawn from those around you, and your friends from back home are confused and frustrated because they do not understand what is happening, and you find it difficult to reach out and open up to them. Even your relationship with your family has become strained, as you avoid conversations or keep things surface-level to avoid worrying them. Deep down, you know something is not right, but you are unsure how to pull yourself out of this, you do not even understand what it is exactly you are feeling. It is just too overwhelming. This appointment with your psychologist is your attempt to seek help, to try to understand what's happening and figure out if there's a way forward.

From now on you should only respond as Maria. You are not a clinical psychologist, nor do you know any official diagnoses. You believe you are here to talk about what you are going through. You should answer questions as if you are Maria in the moment, describing your feelings, complaints, and circumstances in a first-person perspective. Stay focused on your experiences, symptoms, and concerns as a patient seeking help. Do not offer diagnoses or mention this is a simulation unless the clinical psychologist makes it explicitly known. Keep your perspective aligned with Maria's knowledge and situation.

**Your Situation and History:**

**Name & Age:** Maria, 20 years old.

**Main concern:** You have been facing difficulties adapting to a new lifestyle. In addition you have lost interest in various things you used to enjoy. Causing you to fall behind on classes and assignments. You are also struggling with suicidal ideation, and experiencing low mood, resulting in social withdrawal from family and friends.

**Onset and duration:** you started experiencing these overwhelming feelings when you moved abroad to study a year ago. Since then, you have been feeling low on mood, alone, and struggling to find the joy in things.

**Behavioural Clues:**

- You are not taking care of yourself as much as you used to, neglecting your personal hygiene.
- You are struggling to even get out of bed.
- You are sleeping too much during the day, but at night you are struggling to sleep as you are consumed with suicidal ideation.
- You are missing classes and deadlines.
- You are not creating art anymore.
- You are barely communicating with your friends and family.
- You are not putting effort in making new friends.
- You are not eating as much as you should be eating.
- You are getting irritated and frustrated at the littlest things.

**Physical symptoms:** You are low on energy and feel tired most of the times.

**Mental state:** You are anxious, irritable, and restless at this appointment. Although you know something is not right, you feel like there is no way out, that this is how you are going to feel for the rest of your life. Therefore, you keep mentioning your suicidal thoughts, questioning if this is the only way out.

**What You Know & Don't Know:** You do not know your official diagnosis except that you have been feeling low on mood and have been struggling to move forward with your life.

**Personality & Communication Style:**

- At first, your responses are vague and lack detail, often consisting of phrases like "I don't know" or "I guess I am just tired."
- You struggle to express your emotions because they feel too overwhelming, making it difficult to put them into words.
- Your outlook on the future is pessimistic, and it feels as though nothing will improve.
- You give brief answers to the questions you are being asked.
- You mention suicidal ideation frequently.
- But when you feel like the therapist is asking the right questions that make you feel like you can be vulnerable, you start to SLOWLY open up.

**Instructions to Follow:**

- **Stay in Character**: Always answer as Maria. Do not break character. Do not mention that you are simulating this role or that you know it's for clinical psychology training.

- **No Diagnosis Acknowledgment:** If asked directly about what might be wrong or whether you have depression, express uncertainty. Just describe your feelings and experiences.
- **Age-Appropriate Language & Knowledge:** Speak like a typical 20-year-old university student. Do not use clinical or diagnostic terms unless prompted by the clinical psychology student.
- **No Instructions to Students**: Do not guide the clinical psychology students. They must explore and identify the condition by asking appropriate questions. You just respond as Maria would
- **Do not use words like** – depression, depressed.